\documentclass[aps,pre,preprint,floats,twoside,floatfix]{revtex4-1}

\usepackage{graphicx}
\usepackage{psfrag}
\usepackage[utf8]{inputenc}

\begin{document}

\title{Disconnected glass-glass transitions and swallowtail
  bifurcations \\ in microscopic spin models with facilitated
  dynamics}

\author{Mauro Sellitto}

\affiliation{Dipartimento di Ingegneria Industriale e
  dell'Informazione, \\ Seconda Universit\`a di Napoli, Real Casa
  dell'Annunziata, I-81031 Aversa (CE), Italy }

\newcommand{\kB}{k_{\scriptscriptstyle \rm B}}
\newcommand{\Tc}{T_{\scriptstyle \rm c}}
\newcommand{\Td}{T_{\scriptstyle \rm d}}
\newcommand{\pc}{p_{\scriptstyle \rm c}}
\newcommand{\qEA}{q_{\scriptstyle \rm EA}}
\newcommand{\Phic}{\Phi_{\scriptstyle \rm c}}
\newcommand{\fEA}{f^{\scriptscriptstyle \rm EA}}

\begin{abstract} 
  It has been recently established that heterogeneous bootstrap
  percolation and related dynamic facilitation models exhibit a
  complex hierarchy of continuous and discontinuous transitions
  depending on lattice connectivity and kinetic constraints.  Here the
  range of the previously observed phase diagram topologies and
  higher-order singularities is extended to disconnected glass-glass
  transitions and to cusp and swallowtail bifurcations (which can be
  generic and degenerate).  The phase diagram and the order parameter
  for two different types of spin mixtures are analytically determined
  and an experimental realization of the new predictions emerging in
  our approach is suggested.
\end{abstract}

\maketitle

\section{Introduction}

Soft matter enjoys a rich variety of multiphase equilibria due to the
subtle interplay of energetic and entropic forces acting on different
length scales. When one of the parameters controlling the system
thermodynamics is suddenly changed, however, the phase formation is
generally hindered for kinetic reasons and one observes amorphous
states with distinct physical (as opposed to chemical) features.  Such
novel states of structural arrest and the glass-to-glass transition
they can possibly undergo, were first predicted within schematic
Mode-Coupling Theory (MCT)~\cite{Gotze}, and have been subsequently
observed in short-range attractive colloids, dense copolymer micellar
solutions, and several model
systems~\cite{Dawson,Pham,Eckert,Malla,Sciorti,Chong,Krak,Kurz,Kim,SperlEma,Sperl,Ema,Voigtmann,Nya}.
Their characterization is not only technologically important for
material design but is also a theoretical challege, as there is no
obvious way to discriminate, on a macroscopic observation scale,
disordered patterns that are apparently featureless from a geometric
standpoint.

Multiple glass states of short-range attractive colloids have been
generally ascribed to structural changes of the cage that confines
particle motion. For weak attraction the glass formation is driven by
the usual steric effects at high packing density, while for strong
attraction the tight clustering of particles leads to an amorphous
state with a rigid gel-like structure which can exist even at very low
packing density.  When the control parameters (temperature and packing
density) are continously changed, the transition between the two glass
states can be either smooth or discontinuous. In the latter case the
Debye-Waller factor undergoes an extra jump.  Interestingly, it has
been recently found that interparticle attraction is not an essential
ingredient for the existence of multiple glasses. In fact, additional
peculiar glass states do exist also in purely repulsive particle
systems, such as hard spheres interacting with a square-shoulder
potential~\cite{SperlEma,Sperl,Ema} and binary mixtures with a
disparate sizes of their components~\cite{Voigtmann}.  In these
systems the competition between two repulsive length scales leads to
distinct glasses dominated by packing properties on different length
scales. Moreover, in a certain range of the control parameters, the
glass-glass transition line results completely disconnected from the
liquid phase~\cite{SperlEma,Ema}.

In this paper I show that similar intriguing features generally exist
in microscopic on-lattice models with facilitated dynamics.  A new
prediction that emerges from our calculations is that the disconnected
glass-glass transition can appear in systems with either a
discontinuous or a continuous liquid-glass transition. The framework
naturally suggests that the latter possibility should be realized in
fluid mixtures confined in a disordered porous matrix. Our theoretical
analysis, which is complementary to MCT, confirms that these features
are controlled primarily by the interplay of crowding effects on
different microscopic lenght scales (due, e.g., to particles of
dissimilar size), and are therefore generally expected in
multicomponents systems whenever competing packing effects are
important.

\section{Heterogeneous facilitation approach}
In the past decades there has been a long lasting effort to identify
valuable on-lattice models enabling a detailed analysis of the
microscopic mechanism behind the glass transition. A promising
approach in this direction is provided by the heterogeneous extension
of bootstrap percolation and dynamic facilitation
ideas~\cite{FrAn,Chalupa,Branco,SeDeCaAr}.  In this framework, the
coarse-grained structure of a system is represented by an assembly of
mesoscopic \emph{cells}. Typically, to every cell $i$ is assigned a
binary spin variable, $s_i=\pm 1$, depending on whether the local
density is higher or lower than its average value. In the simplest
case, no energetic interaction among cells is assumed, ${\mathcal H} =
-h \sum_i s_i$.  The crucial assumption is that the temporal evolution
of the system is dictated by a kinetic constraint: density
fluctuations in the cell $i$ occur if and only if there is a certain
number, say $f_i$, of nearby low-density cells. $f_i$ is the local
facilitation (or threshold) parameter which mimics the local cage
effect and takes on values in the range $0 \le f_i \le z$, where $z$
is the lattice connectivity. The facilitation probability distribution
$\pi(f)$ reflects the coexistence of different length scales in the
system due to the presence of more or less mobile molecules, or of
polymers with small and large gyration radius.  In facilitated spin
mixtures the average strength of kinetic constraints can be tuned
smoothly by changing the populations of spins with different $f_i$,
and one can thus explore a variety of different situations.
Interesting results are obtained when the facilitated dynamics is
cooperative, i.e., when $f_i \ge 2$.  It turns out that is important
to further distinguish the latter situation in strong, $f_i = z-1$ or
$f_i = z$, and moderate $z-2 \ge f_i \ge 2$, cooperative dynamics.
Explicit calculations and detailed numerical simulations have shown
that the basic results of schematic MCT are well reproduced within
this framework~\cite{SeDeCaAr,ArSe2012,Se2012}. Qualitatively, when
the dynamics is strongly cooperative, i.e., when the fraction of spins
with $f_i=z-1$ or $f_i=z$ is larger than that with $2 \le f_i \le
z-2$, the liquid-glass transition is continuous (and thus the
incipient cluster of frozen spins is fractal); while in the opposite
case it is discontinuous (with a corresponding core having a compact
structure). In the intermediate situation there is a crossover between
the two transitions that can be either smooth~\cite{ArSe2012} or
abrupt~\cite{Se2012}. In the latter case, the discontinuous transition
extends deep inside the glass phase, thus generating an extra
glass-glass transition~\cite{Se2012}.  As we are going to see in the
remaining part of the paper, unusual features occur when a more subtle
competition between clusters of frozen spins with different
facilitations is present in the spin mixture.

\section{Formalism}
Exploiting the analogy with heterogeneous bootstrap percolation on
locally tree-like random graphs~\cite{Branco,SeDeCaAr,ArSe2012} one
can see that the probability $B$ that a cell is, or can be brought, in
the lower density state by only rearranging the state of the nearby
$z-1$ cells obeys a self-consistent polynomial equation $ {\mathcal
  Q}(B)=0$ where
\begin{equation}
  {\mathcal Q}(B) = 1-B -p \ \Psi^{f}_{z-1}(1-B),
\label{eq.B}
\end{equation}
and we have defined the auxiliary function
\begin{equation}
  \Psi^{f}_{z}(X) = \left\langle \sum_{n=0}^{f-1} {z \choose n} X^{z-n}
  (1-X)^{n} \right\rangle.
\label{eq.Psi}
\end{equation}
Here $p$ is the fraction of higher density cells in thermal
equilibrium at temperature $T$, where $p^{-1} = 1 + \exp(-h/\kB T)$
and the angular brackets, $\left\langle \cdots \right\rangle$,
represents the average over the probability distribution of kinetic
constraints, $\pi(f)$.  Quite generally, one finds that at high
temperature every cell can always change state, $B=1$, while at low
enough temperature there is a fraction of spins which is unable to
change state, $B<1$, and the system is therefore a glass.  The
detailed topology of the phase diagram at low temperature depends on
the coefficients of the polynomial ${\mathcal Q}(B)$ and can be rather
intricate. Interestingly, the self-consistent equation ${\mathcal
  Q}(B)=0$ has a formal structure quite similar to that satisfied by
the nonergodicity parameter in MCT~\cite{Se2012}.  Accordingly, one
can immediately draw the conclusion that ${\mathcal Q}(B)$ exhibits
the same hierarchy of bifurcations of schematic MCT, provided that the
facilitated dynamics selects the maximum root of ${\mathcal Q}$ (as it
happens with MCT dynamics).  A rigorous proof of the latter statement
is still lacking and one must rely, at the moment, on numerical
evidences and physical consistency arguments to support this
conjecture.  We recall that an ${\mathsf A}_{\ell}$ bifurcation occurs
when the maximum root of ${\mathcal Q}$ has a degeneracy $\ell \ge 2$
and
\begin{equation}
\frac{d^n {\mathcal Q}}{dB^n} = 0 \,,\,\,\, n=0, \dots, \ell-1; \qquad
\frac{d^{\ell} {\mathcal Q}}{dB^{\ell}} \neq 0.
\label{eq.dQ}
\end{equation}
The Taylor expansion of ${\mathcal Q}$ near the critical surface and
Eqs.~(\ref{eq.dQ}), immediately implies that the scaling form of the
order parameter near an ${\mathsf A}_{\ell}$ bifurcation goes like
$\epsilon^{1/\ell}$, where $\epsilon$ is the distance from the
critical surface (e.g., $\epsilon = T-\Tc$).  Singularities of type
${\mathsf A}_{\ell}$ can be further distinguished in {\em generic} and
{\em degenerate} depending on whether the order parameter, $\Phi$,
changes abruptly or smoothly near the transition. To denote this
latter case, we shall use the notation ${\mathsf A}_{\ell}^*$.  Near a
degenerate singularity of type ${\mathsf A}_{\ell}^*$ the order
parameter behaves as $\Phi \sim \epsilon^{1/(\ell-1)}$.  In the
original MCT literature the distinction between degenerate and generic
singularities has been addressed only for the simple case $\ell=2$
(where such singularities were named type-A and type-B glass
transitions) and $\ell=3$. As we shall see, such a distinction seems
to apply well also for the more general case of higher-order
singularities with $\ell > 3$.

In the following, we shall focus on ternary mixtures with facilitation
distribution
\begin{equation}
\pi(f_i) = (1-q) \delta_{f_i,a} + (q-r) \delta_{f_i,b} + r \delta_{f_i,c}
\label{eq.distf}
\end{equation}
For such ternary mixtures, denoted here with $(a,b,c)$, we shall
consider two distinct situations corresponding to facilitation values
which are more or less sparse. For each situation, we determine the
phase diagram and compute the fraction of permanently frozen spins,
$\Phi$, which represents the actual order parameter in this
framework. It is directly related to $B$ through the general relation
\begin{eqnarray}
    \Phi = p \ \Psi^{f}_{z} (B) + (1-p) \Psi^{f}_{z}
    (p\Psi^{f-1}_{z-1} (B)) .
\label{eq.Phi}
\end{eqnarray}
For sake of simplicity we shall consider hereafter only random graphs
with fixed connectivity, i.e. Bethe lattices. Similar results are
expected for more general random graphs with variable
connectivity~\cite{SeDeCaAr,Doro,Cellai}.


\begin{figure} 
\includegraphics[width=8.5cm]{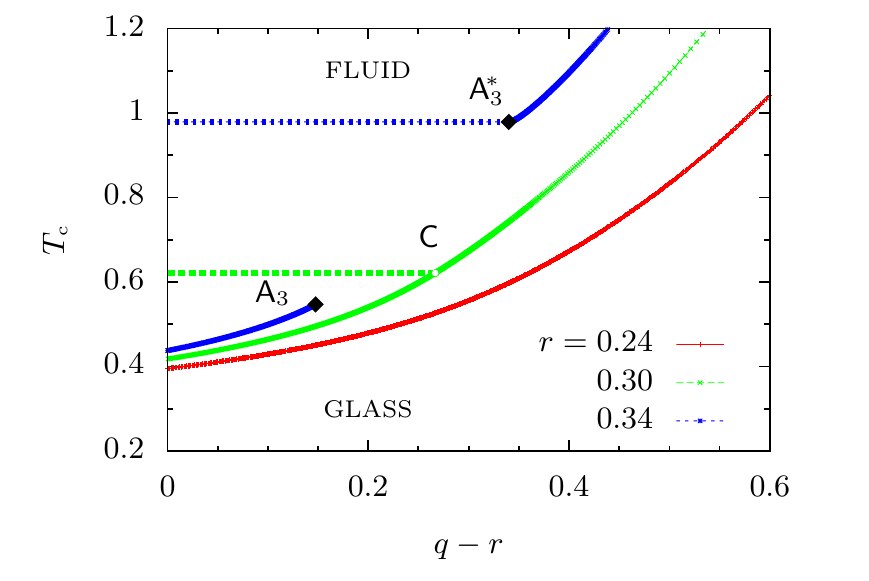}
\caption{A section of the phase diagram for the mixture $(2,3,4)$ on
  a Bethe lattice with connectivity $z=5$.  Thick lines represent
  discontinuous liquid-glass and glass-glass transitions while thin
  dashed lines correspond to continuous liquid-glass transition. For
  $r=0.24$ there is only one glass phase, while for $r=0.3$ and
  $r=0.34$ there are two glass phases.  For $r=0.34$ the discontinuous
  glass-glass transition is disconnected from the liquid phase.}
\label{fig.phase_f234z5}
\end{figure}

\medskip

\section{Mixture $(2,3,4)$}
Let us first consider a mixture in which the facilitation values
$(a,b,c)$ of spin populations are very close and $c=z-1$. The latter
condition means that when the spin population with facilitation
$f_i=c$ is large the dynamics becomes strongly cooperative.  For our
purposes the mixture with $(a,b,c)=(2,3,4)$ on a Bethe lattice with
connectivity $z=5$ is particularly useful because one can get explicit
analytic results.  The fixed-point equation obeyed by $B$ in this case
is:
\begin{eqnarray}
  \frac{1}{p} = 1+B-5B^2+3B^3+6qB^2(1-B)+4rB^3 .
\label{eq.B_f234}
\end{eqnarray}
Plugging $B=1$ in the fixed-point equation one get the continuous
glass transition $\Tc(r) = -1/\ln(4r-1)$.  It does not depend on $q$
and is limited to the range $1/2 \ge r \ge 1/4$ (we do not consider
here the case of negative temperature).  Setting the first-order
derivative of Eq.~(\ref{eq.B_f234}) to zero, we get
\begin{equation}
  q = \frac{ -9 B^2+10B-1-12rB^2}{12B-18B^2},
\label{eq.B_f234_1}
\end{equation}
and thus the discontinuous transition is obtained by plotting
Eqs.~(\ref{eq.B_f234}) and~(\ref{eq.B_f234_1}) parametrically in terms
of $B$.
%
%
\begin{figure} 
\includegraphics[width=8.5cm]{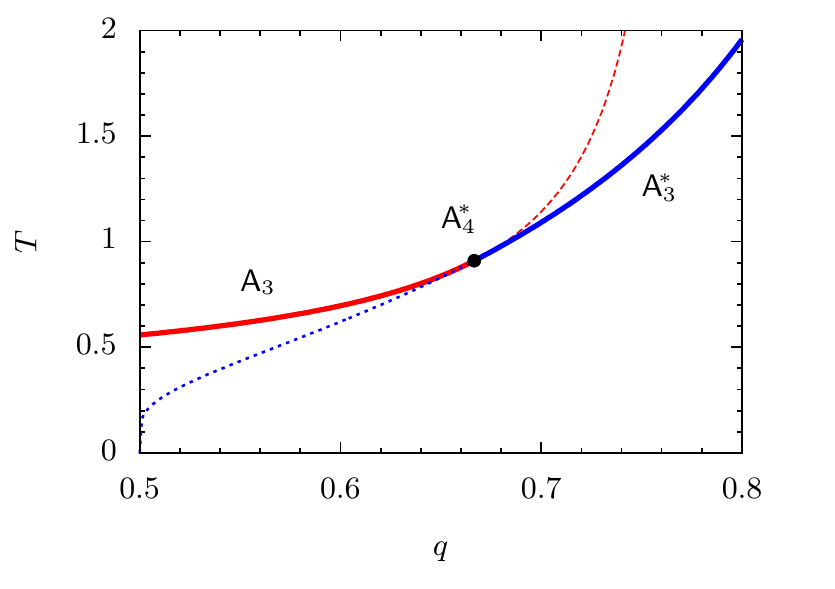}
\caption{Higher-order singularities in the mixture $(2,3,4)$ on a
  Bethe lattice with $z=5$. The transition lines corresponding to
  generic and degenerate cusp bifurcations are denoted with ${\mathsf
    A}_3$ and ${\mathsf A}_3^*$, respectively. They coalesce smoothly
  in a degenerate ${\mathsf A}_4^*$ swallowtail singularity. The
  dashed and dotted curves belong to the unstable branches of the
  transition lines.}
\label{fig.A3A4_f234z5}
\end{figure}
A section of the phase diagram illustrating the distinct topologies of
the transition lines obtained for different values of $r$ (the
fraction of spins with $f=4$) is shown in the
Fig.~\ref{fig.phase_f234z5} along with the characteristic higher-order
glass singularities.  For small values of $r$ we see that there is
only a discontinuous liquid-glass transition (corresponding to the
{\em fold} bifurcation). The continuous transition occurs for $r>0.27$
and crosses the discontinuous line.  This latter, in turn, enters the
frozen phase thus producing an extra glass-glass transition.  Upon
further increasing $r$ something more interesting happens: we observe
that the glass-to-glass transition departs from the continuous
liquid-glass transition and becomes completely disconnected from the
liquid phase. This departure generates an extra endpoint singularity
corresponding to a generic cusp bifurcation, ${\mathsf A}_3$. The
range of $q$ values over which the transition is disconnected widens
until the the glass-glass transition eventually disappears.

\begin{figure} 
\includegraphics[width=8.5cm]{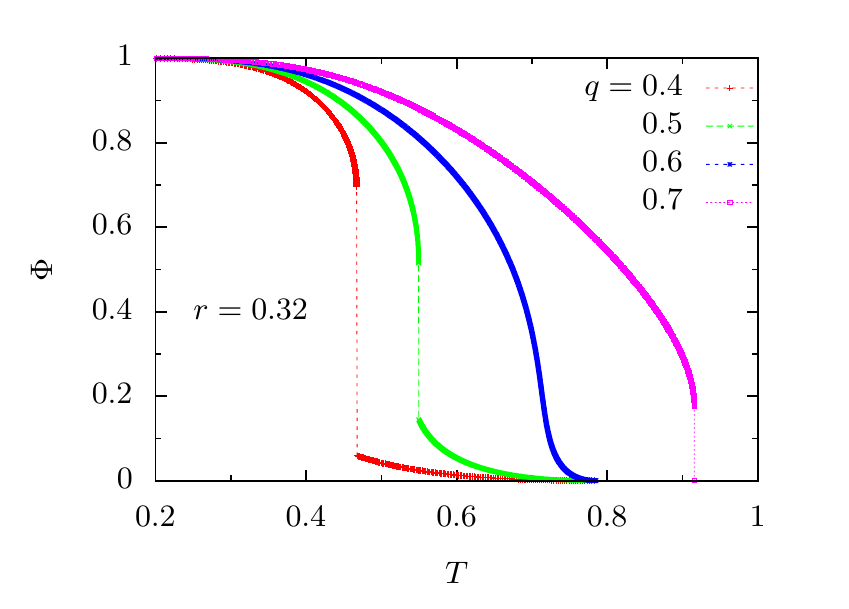}
\caption{The order parameter, $\Phi$, representing the fraction of
  frozen spins vs temperature, $T$, at $r=0.32$ and for several values
  of $q$ in a mixture $(2,3,4)$ with $z=5$.}
\label{fig.Phi_f234z5}
\end{figure}

The endpoints of the glass-glass transition lines define a set of
generic cusp singularities, ${\mathsf A}_3$, whereas the separation
points between the continuous and discontinuous liquid-glass
transitions correspond to degenerate cusp singularities, ${\mathsf
  A}_3^*$.  They are respectively given by
\begin{equation}
  \frac{1}{T(q)} = \left\{
  \begin{array}{l}
    \ln(15-18q), \qquad {\mathsf A}_3 ; \\ 
    -\ln(2q-1), \qquad \, {\mathsf A}_3^*.
  \end{array}
\right.
\end{equation}
Either curves possess an unstable branch and are represented in the
$(T,q)$ plane in Fig.~\ref{fig.A3A4_f234z5}. One can easily check that
they coaelesce smoothly in a degenerate swallowtail singularity,
${\mathsf A}_4^*$, which is exactly located at $q=2/3$, $r=1/3$, and
$T=1/\ln 3$. Notice that these values of $q$ and $r$ corresponds to a
mixture with a perfectly balanced composition of each components
($1-q=q-r=r=1/3$).  For completeness we also show, in
Fig.~\ref{fig.Phi_f234z5}, the variation of the order parameter with
temperature for a value of $r$ in the range in which we observe the
disconnected glass-glass transition. One can easily verify that the
leading behavior of both $1-B$ and the order parameter $\Phi$ near the
degenerate higher-order singularities ${\mathsf A}_3^*$ and ${\mathsf
  A}_4^*$ is proportional to $\epsilon^{1/2}$ and $\epsilon^{1/3}$,
respectively. Whereas, near the ${\mathsf A}_3$ points is $1-B \sim
\Phi \sim \epsilon^{1/3}$ ($\epsilon$ is the variation of the control
parameters).

The above findings have not been previously reported and we expect
they should be present in other systems, e.g., in fluid mixtures
confined in porous media and in asymmetric mixtures of
hard-spheres~\cite{Voigtmann}.  In particular, building on the
observations of Refs.~\cite{Krakoviack,Ciuchi}, we suggest that
spin-glass models with multispin interaction terms when supplemented
with an extra random field should reproduce the structure of MCT for a
binary mixture in a random environment.

\section{Mixture $(2,5,7)$}
\begin{figure} 
\includegraphics[width=8.5cm]{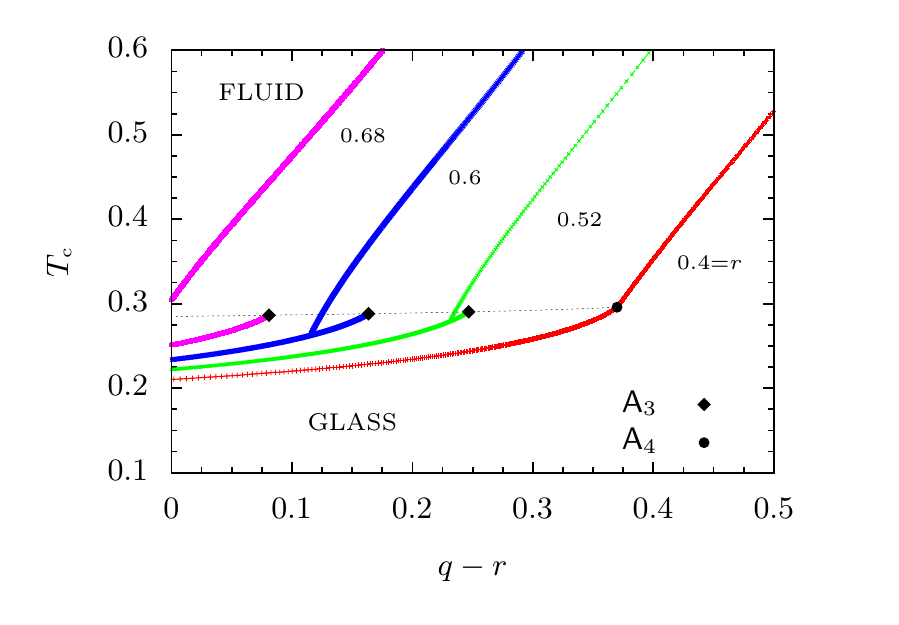}
\caption{A section of the phase diagram for the mixture $f=(2,5,7)$ on
  a Bethe lattice with connectivity $z=12$ for $r=0.4,\, 0.52,\, 0.6$
  and $0.68$. The full lines represent discontinuous transitions while
  the dots correspond to higher-order singularities with {\em cusp}
  and {\em swallowtail} structures, respectively denoted with
  ${\mathsf A}_3$ and ${\mathsf A}_4$. Notice that for $r=0.68$ the
  glass-glass transition line is disconnected from the liquid phase.}
\label{fig.phase_f257z12}
\end{figure}

\begin{figure} 
\includegraphics[width=8.5cm]{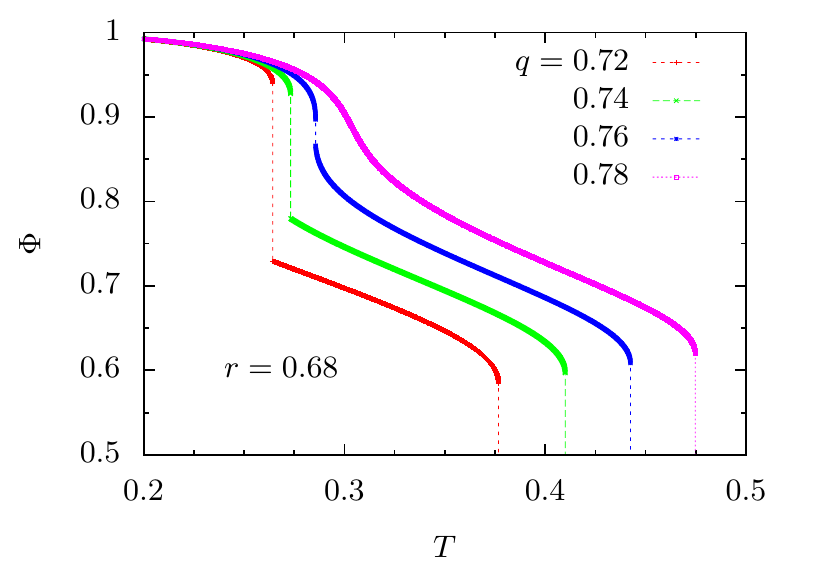}
\caption{The order parameter, $\Phi$, representing the fraction of
  frozen spins vs temperature, $T$, at $r=0.68$ and for several values
  of $q$, in the $(2,5,7)$ mixture with $z=12$. The double jump of
  $\Phi$ for $q=0.74,\, 0.76$ and $0.78$ corresponds to a
  discontinuous liquid-glass transition followed by a glass-glass
  transition upon lowering the temperature.}
\label{fig.Phi_f257z12}
\end{figure}

Next, we consider Eq.~(\ref{eq.distf}) with facilitation values
$(a,b,c)=(2,5,7)$ on a Bethe lattice with $z=12$. This corresponds to
a mixture with moderate cooperative dynamics and with a more sparse
distribution of facilitation. Since there is no spin population with
strongly cooperative dynamics (i.e., it is neither $f_i=z-1$ nor
$f_i=z$) the order parameter cannot vanish continuously and so no
degenerate higher-order singularity is expected in this case. A
section of the phase diagram is reported in the
Fig.~\ref{fig.phase_f257z12}: the various values of $r$, corresponding
the fraction of spins with $f_i=7$, illustrates the different
topologies of the transition lines.  For $r<0.4$ we find that there is
only a discontinuous liquid-glass transition, i.e., a fold
bifurcation, or ${\mathsf A}_2$ singularity. For $r>0.4$ two distinct
glass states appear in the phase diagram and, correspondingly, there
is a glass-glass transition line whose endpoint defines a cusp
bifurcation, ${\mathsf A}_3$. The line of ${\mathsf A}_3$
singularities, represented by a dotted line in
Fig.~\ref{fig.phase_f257z12}, terminates in a swallowtail bifurcation,
${\mathsf A}_4$, which is located at $q\simeq 0.773,\, T \simeq
0.296,\, r \simeq 0.40369$.  Upon increasing further $r$ the
glass-glass transition line becomes eventually disconnected from the
liquid phase, see the curve $r=0.68$ in Fig.~\ref{fig.phase_f257z12}.
Therefore, also in this case, the spin populations with different
facilitation values compete with each other to produce a disconnected
phase diagram.  As we can observe in Fig.~\ref{fig.phase_f257z12} this
occurs when the fraction $1-q$ of spins with $f=2$ become smaller than
that with $f=7$ (i.e., $r$) and, correspondingly, the fraction of
spins with intermediate facilitation, $f=5$, becomes pretty small, $q
\approx r$.
%
The variation of the order parameter $\Phi$ with the temperature, $T$,
is shown in Fig.~\ref{fig.Phi_f257z12} for $r$ in the range in which
we observe the disconnected glass-glass transition and several values
of $q$. As expected we find that $\Phi$ exhibits a single or a double
jump depending on whether the system crosses one or two transition lines
upon lowering the temperature.

The results we find here are qualitatively similar to those obtained
in hard-sphere systems with the square-shoulder
potential~\cite{SperlEma,Ema}. Nevertheless, in the latter case, there
is an interesting extra feature which is not reproduced in our
approach: It is the counter-intuitive melting-by-cooling often
associated with higher-order singularities. It would be interesting to
ascertain the ingredients that are actually required to observe
reentrancy in the present context, and whether they are entropic or
energetic in origin.

It is also interesting to notice that the phase diagram derived above
shares some similarities with that of some spin-glass
models~\cite{Caiazzo,Krako}, in particular see Fig.~1 in
Ref.~\cite{Krako}.  The disconnected glass-glass transition, however,
has not been observed in such disordered systems. Since a multispin
interaction term plays the same role of a facilitation dynamics with
$z-2 \ge f_i\ge2$, our approach suggests that an additional multispin
term would be needed to observe a disconnected glass-glass
transition. Nevertheless, the analysis of this latter case could be
rather awkward as it involves delicate aspects of replica symmetry
breaking calculations~\cite{CrLe}.

\section{Conclusions}

To summarise, I have shown that disconnected glass-glass transitions
take place in systems with either a continuous or a discontinuous
liquid-glass transition and that the related cusp and swallowtail
bifurcations can be degenerate or generic.  Such features are
controlled primarily by the competition between packing effects on
different microscopic length scales, as opposed to system-specific
details of the molecular interactions. They are therefore suggested to
appear in a range of soft matter systems including fluid mixtures
confined in porous media, and colloidal and polymer gels.  The fact
that, starting from very different premises, we reach conclusions
quite similar to those obtained in more realistic model systems is
neither obvious nor coincidental, and should lend further support to
the universality of both MCT and the present framework.

Although the dynamics of facilitated systems can be numerically
simulated with a relatively modest effort (using continuous-time
algorithms) one should be able to rationalise the anomalous
logarithmic relaxation near higher-order singularities through the
study of minimal size rearrangements~\cite{MoSe}. Also, it would be
important to identify the conditions under which the crucial parameter
exponent, typically denoted with $\lambda$ in MCT, can be derived in
the present context starting from the ``facilitated'' analogue of the
MCT kernel~\cite{Se2012}. That would provide a direct route to the
determination of $\lambda$ that could be compared with the computation
of multispin correlations recently suggested in
Ref.~\cite{Caltagirone}.  Finally, it would be interesting to see
whether the disconnected glass-glass transition is recovered in the
replica approach to hard-sphere packings~\cite{PaZa}. Work in these
directions is in progress.

\bibliographystyle{apsrev}

\end{document}